\documentclass[prb,reprint,floatfix]{revtex4-1}

\usepackage{amsmath}  
\usepackage{amsfonts} 
\usepackage[dvips]{epsfig}

\begin{document}

\title{Bridging the Gap between Crystal Theory and Semiconductor Physics}

\author{William R. Frensley}
\email{frensley@utdallas.edu} 

\affiliation{Department of Electrical and Computer Engineering, University of Texas at Dallas, Richardson, TX  75080}

\date{\today}

\begin{abstract}
 The theory of perfect crystals, founded upon the Bloch theorem, gives an understanding of extended quantum states grouped into energy bands,
and permits the derivation of the dynamics of electrons in those states.  The semiconductor physics used to explain the operation of electronic devices
treats the (imperfect) semiconductor crystal as a uniform effective medium in which positively and negatively charged quasiparticles mostly obey Newtonian
dynamics, and in which the chemistry of impurity atoms is far different from that of those same atoms in free space.  The connection between these two pictures can 
be made by made by invoking a mathematical transformation that takes the finite-temperature, impure device structure and algebraically subtracts from it
a perfect crystal, leaving only the residual differences to be analyzed.  This notion of the residual difference offers a conceptual basis for understanding many 
aspects of semiconductor physics, including the properties of impurity states and heterogeneous interfaces.  The mesoscopic transformation that underlies the 
residual-difference picture provides the systematic way to define a concept that
is essential to the understanding of semiconductor devices: a position-dependent band structure. 
\end{abstract}

\maketitle

\section{Introduction} 

The physics of semiconductor devices is taught in a number of contexts, most prominently in courses dedicated to this topic in Electrical Engineering curricula, but also in 
Modern Physics courses (particularly when an associated laboratory includes measurements of the Hall effect), and other courses in solid-state physics and materials science.
Semiconductor devices are also the prototypical nanostructured systems of technological significance, and an understanding of how this field handles the connections between different levels of
spatial-scale abstraction can benefit those who wish to pursue work in nanotechnology.

The traditional expositions of semiconductor physics make a rather awkward leap from the microscopic physics of perfect crystals to the meso- or macroscopic picture in which all subsequent
discussion is grounded.  In the microscopic picture we treat only perfect, unbounded crystals to which the Bloch theorem applies.  We make the assumption that we have access to 
a sufficiently complete solution to the electronic structure problem posed by the crystal, and thus can know the energy-band structure.  The band structure consists of the dispersion
relations $E_b(\bf k)$, the energy in band $b$ of the state with wavector $\bf k$, which is assumed to lie within the first Brillouin zone, and also the associated wavefunctions, in
some useable form.  From this energy-band structure we can derive the laws of dynamics for band electrons, which consist of the group velocity theorem:\cite{Kroemer1975}
\begin{equation}  \label{eqn:groupVelocity}
 {\bf v} = \frac{1}{\hbar} \nabla_{\bf k} E_b(\bf k) ,
\end{equation}
and the wavevector-acceleration theorem:\cite{Kroemer1986}
\begin{equation} \label{eqn:kAcceleration}
 \frac{d{\bf k}}{dt} = \frac{\bf F}{\hbar},
\end{equation}
where $\bf F$ will typically be the Lorentz force.

The macroscopic picture that is supposed to follow from this foundation is that of a semiconductor as a uniform effective medium, from which all trace of the underlying crystal structure
has disappeared.  Within this effective medium discrete Newtonian quasiparticles move freely, the quasiparticles being negatively-charged electrons and positively-charged holes.  Each of 
these types of quasi-particle moves as if it has an ``effective mass'' $m^*$ that is typically of the order of $0.1m_0$, $m_0$ being the mass of the electron in free space.  
Impurity atoms can be embedded within this medium by technological processess, and once they are in place they generally act as spatially fixed charges of unit magnitude and either sign.
The semiconductor is also a dielectric with relative permittivity $K$ which is typically a bit greater than 10.  From this picture the development of the subject proceeds to the equilibrium
distributions of electrons and holes, and their nonequilibrium interactions including electron-hole pair generation and recombination, and transport in response to applied electric and magnetic
fields and concentration gradients.  These last elements are the basis for describing the operation of diodes, transistors, and other semiconductor devices.

\section{The Mesoscopic Transformation}
There is an intermediate step in the logical development that is the key to bridging the gap described above. It is generally described as ``effective-mass thoery,'' and the seminal works 
are those of Slater\cite{Slater1949} and of Luttinger and Kohn\cite{Luttinger1955}.  In the present work we will consider these to be particular implementations of a broader category
of \emph{mesoscopic transformations}.  Effective-mass theory is generally presented as an approach to treating a system consisting of a periodic potential $V_\text{periodic}$ and an 
``external'' slowly-varying potential $V_\text{ext}$.  It is generally assumed that one unique $V_\text{periodic}$ will adequately determine the energies and wavefunctions of all the 
states of interest, typically those of the valence bands and lower conduction bands.  By factoring out a basis function that approximates the behavior of the microscopic
wavefunction within the crystal unit cell, the rest of the wavefunction becomes a slowly-varying \emph{mesoscopic wavefunction} $\Psi_b$ (often called the envelope function when the Luttinger-Kohn
formulation is used).  The mesoscopic wavefunction is always associated with a particluar band $b$, hence the need to include a band index in the notation.

The mesoscopic wavefunction obeys the effective-mass Schr\"odinger equation which is of the general form:
\begin{equation}
 i\hbar\frac{\partial\Psi_b}{\partial t} = E_b(-i\nabla) \Psi_b + V_\text{ext} \Psi_b,
\end{equation}
where $E_b$ is here the functional form of the dispersion relation applied to the gradient operator.  We usually expand this function about its extremum (assumed for the present to be located at
${\bf k} = 0$), yielding:
\begin{equation*}
 E_b({\bf k}) \approx E_b(0) + \frac{\hbar^2 k^2}{2m^*},
\end{equation*}
and we now see the origin of the effective mass.  The effective-mass Schr\"odinger equation now takes a more familiar form:
\begin{equation} \label{eqn:effMassSchEq}
  i\hbar\frac{\partial\Psi_b}{\partial t} = - \frac{\hbar^2 }{2m^*} \nabla^2 \Psi_b + \left(E_b + V_\text{ext} \right) \Psi_b .
\end{equation}
There is a distinct equation for each energy band of interest.  Note that the periodic potential remains in the problem only through the dispersion relations $E_b(\bf k)$.

Before we examine the conceptual consequences of this absorption of the periodic potential we need to look more clearly at what is actually contained in that potential.
First, we will assume that the ideal crystal bandstructure computation used the pseudopotential approach.  The psuedopotential is an effective potential that is much weaker than
the realistic potential within an atom, which is constructed to allow no bound core states, but to produce accurate wavefunction energies for states with energies well above those 
of the tightly-bound core electron states.  The reason for insisting on the use of pseudopotentials (apart from the fact that all modern electronic-structure computations use this 
approach) is that this allows us to assert that the ion core (pseudo)potentials for different elements from the same column of the periodic table differ by only small amounts, even 
though their nuclear charge and core electron configurations differ a great deal.  In addition to the core pseudopotentials, $V_\text{periodic}$ will include contributions from the 
Hartree potential, which must be self-consistent with the valence-electron charge
distribution, and the appropriate exchange-correlation potential.  

Now we can articulate a more general interpretation of the mesoscopic transformation than merely cancelling out the periodic potential.  When we apply this transformation, we will not
only cancel out the periodic potential, we will cancel, or subtract, every quantity with which the potential is self-consistent.  \emph{The mesoscopic transformation algebraically 
subtracts an ideal crystal from the non-ideal system that is of interest.  This subtraction includes the ion-core potentials, the valence-electron charge distribution and the 
occupation probabilities for each electron state in the system.}  And, as was the original intent of the approach, the strong periodic crystal potential is subtracted out, leaving
only the residual Hartree potential due to the charges that are present in the residual differences that define the system of interest.  
There are two other consequences of this transformation: The residual charge carriers (electrons and holes) obey the effective-mass Schr\"odinger equation of the appropriate energy band,
and electrostatic interactions within the residual system must be computed using the dielectric constant of the semiconductor material.  The last requirement follows from the fact that
we may be able to abstract away the valence charge distribution, but we cannot abstract away its self-consistent response to applied electric fields.  With this in mind, we can specify
that $V_\text{ext}$ should be the Hartree potential derived from the {\it residual} system by solving Poisson's equation in the dielectric medium for the residual charge distribution, 
subject to bundary conditions derived from state of any electrical contacts to the system.  (The particle densities in the residual system are sufficiently dilute that exchange interactions
are generally negligible.)

\begin{figure}[t!b!]
\centering
\epsfig{file=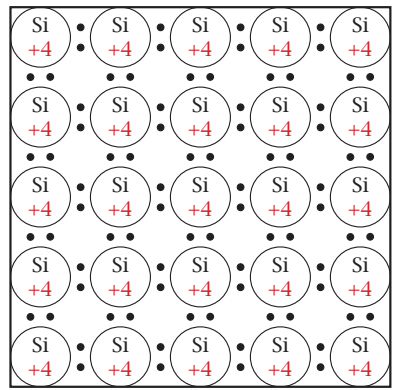, width=2.0in}
\caption{Schematic drawing of a perfect silicon crystal, illustrating the ion cores and the valence electrons.}
\label{fig:MesoPerfCrys}
\end{figure}

\begin{figure}[t!b!]
\centering
\epsfig{file=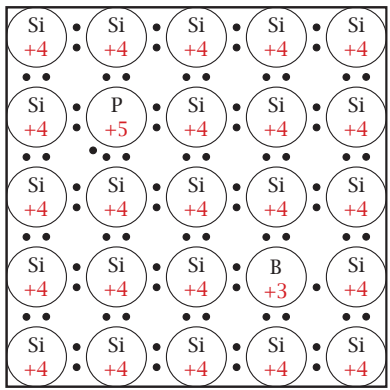, width=2.0in}
\caption{Schematic of a crystal with impurity atoms introduced as is common in device fabrication processes.}
\label{fig:MesoImpCrys}
\end{figure}

\begin{figure}[t!b!]
\centering
\epsfig{file=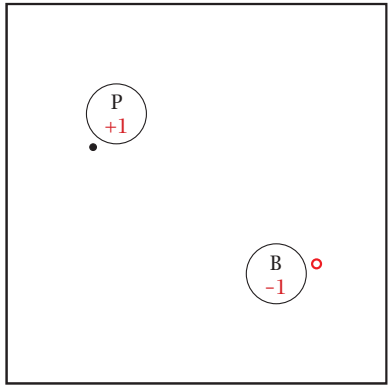, width=2.0in}
\caption{Schematic of the simplified system resulting from the mesoscopic transformation which algebraically subtracted Fig.\ \ref{fig:MesoPerfCrys} from 
Fig.\ \ref{fig:MesoImpCrys}.  The open circle where an electron would have been in the perfect crystal represents the hole quasiparticle.}
\label{fig:MesoDifference}
\end{figure}

The workings of the mesoscopic transformation may be visually illustated using three Figures.  Fig.\ \ref{fig:MesoPerfCrys} schematically illustrates a perfect silicon crystal.
Fig.\ \ref{fig:MesoImpCrys} shows a Si crystal in which impurity atoms have been introduced.
Finally, Fig.\ \ref{fig:MesoDifference} shows the enormous simplification that is evident in the residual system after the algebraic subtraction of the ideal crystal.

\section{Occupation Factors and quasiparticles}

Expressing the effect of the mesoscopic transformation as a subtraction operation clarifies a number of points.  Most importantly it explains why quantities that are 
manifestly non-negative like particle spatial densities and quantum state occupation probabilities can now take negative values.  In particular there can be negative
occupation probabilities for states in the valence band, otherwise known as a positive density of holes.  The dynamic properties of a hole are indistinguishable from those
of a positvely-charged particle with positive mass, particularly in measurements of the Hall effect, and therefore we can regard it as a quasiparticle with those properties.  
The point that is essential when one attempts to explain how the absence of a negatively-charged electron can move like a positively-charged particle is that the state from
which the electron is missing has a negative effective mass.  So the short explanation is that the absence of a negatively-charged, negative effective-mass particle is equivalent 
to the presence of a positively-charged, positive mass particle.

\begin{figure}[t!b!]
\centering
\epsfig{file=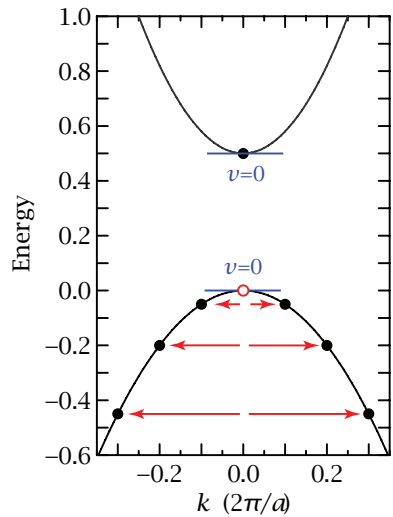, width=2.0in}
\caption{Energy-band structure showing the behavior of the electron and hole quasiparticle.}
\label{fig:HolesFig1}
\end{figure}

But it is often useful to expound upon this point by making a more detailed examination of the dynamical observables, including energy, momentum and charge current.  This 
explanation can be a bit tricky, and it is very easy to draw erroneous conclusions that appear to contradict the positive quasiparticle picture.  Thus it is worthwhile to document
the details of this analysis.  We begin by defining a reference state of the crystal in Fig.\ \ref{fig:HolesFig1}, that is in particlular required for the evaluation the kinetic energies
of the quasiparticles.
In this state there is one extra electron in the conduction band and one missing electron (hole) in the valence band, both assumed to be at ${\bf k} = 0$ in this initial condition.
In what follows, one must bear in mind that the particle velocity is the group velocity (\ref{eqn:groupVelocity}) and that the charge current is in reality carried by electrons and is
defined by:
\begin{equation}  \label{eqn:chgCurrent}
 {\bf j} = - q {\bf v}.
\end{equation}

We quickly see that the conduction-band electron in Fig.\ \ref{fig:HolesFig1} has zero momentum, velocity, and current.  For the valence band, however, it is still the electrons
that contribute the momentum and current.  Thus, we need to sum these quantities over all of the occupied valence-band states.  Thes sums are greatly simplified by the Kramers
degeneracy of $E_b({\bf k})$, which guarantees that the dynamical contributions of a state $|{{\bf k}, b}\rangle$ will be cancelled out by the contributions
of state $|{-{\bf k}, b}\rangle$, provided that both are occupied.  This cancellation is represented in Fig.\ \ref{fig:HolesFig1} by opposing arrows.  Taking this cancellation into
account, we readily see that the valence-band system of Fig.\ \ref{fig:HolesFig1} has zero momentum and current.  The unoccupied hole state also has a zero velocity due to its position at an extremum
of the band structure.

\begin{figure}[t!b!]
\centering
\epsfig{file=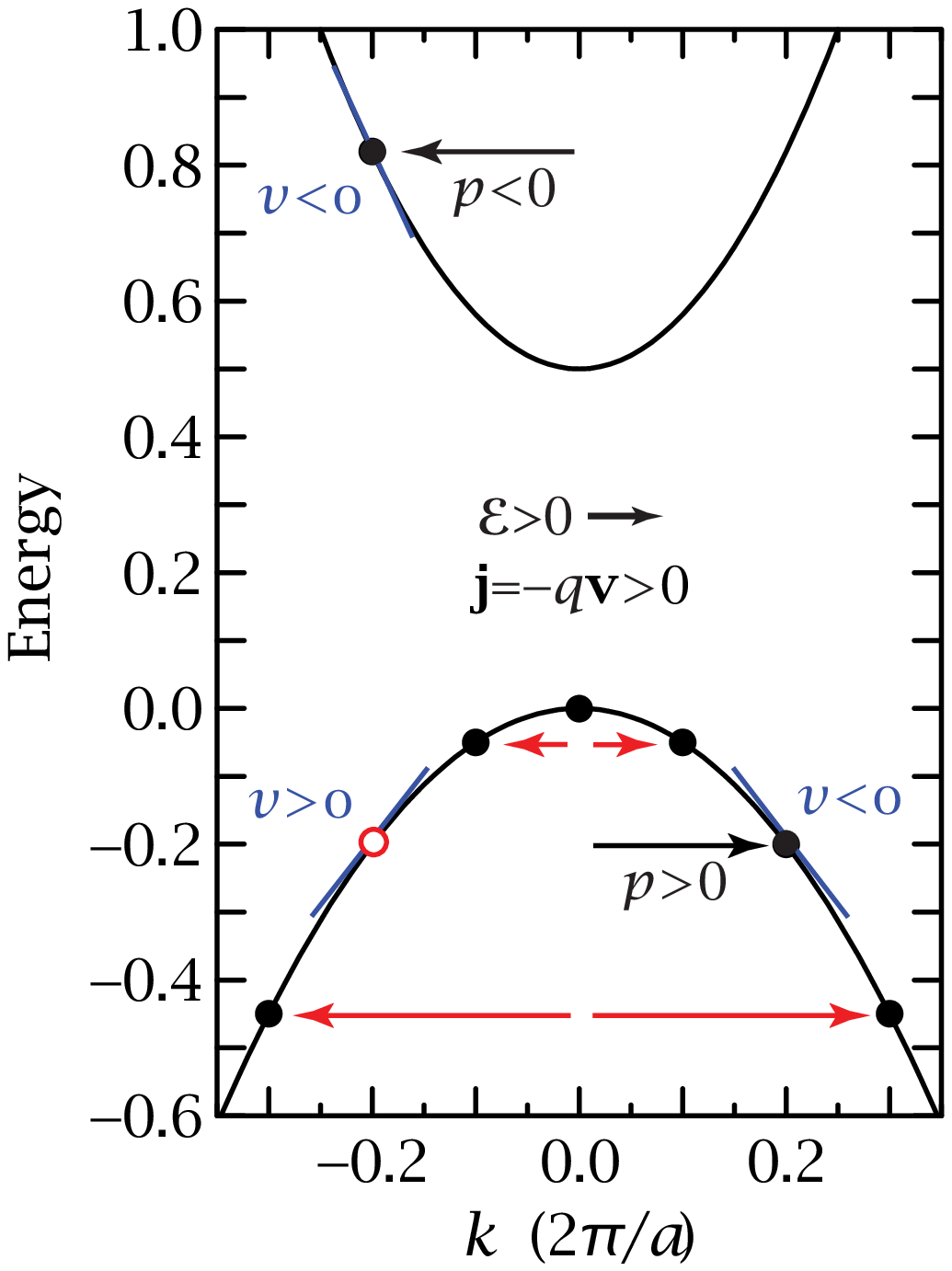, width=2.0in}
\caption{Energy-band structure showing the behavior of the electron and hole quasiparticle.}
\label{fig:HolesFig2}
\end{figure}
We now assume that an electric field $\mathcal{E}$ in the positive direction is applied for a time sufficient to accelerate all the electron states so that their wavevector changes by 
$\Delta {\bf k} = -0.2 (2\pi/a)$, producing the situation shown in Fig.\ \ref{fig:HolesFig2}.  Again, the behavior of the electron in the conduction band is straightforward:
it has acquired a negative momentum indicated by the arrow, a negative velocity indicated by the negative slope of $E_c({\bf k})$, and a positive current density as the product
of the negative charge and negative velocity.  Also, the total energy of the conduction-band system increased by $p^2/2m^*_c$, which we interpret as the electron's kinetic 
energy.

The valence-band system has acquired a positive momentum, indicated by the dark arrow, due to the state at ${\bf k} = +0.2$, whose momentum is not cancelled out.  This state
also contributes the only uncancelled current, which is positive from the product of the negative charge and negative velocity (slope of the valence-band curve).  Note that the
hole state has a positive velocity due to the slope at its position.  The system consisting of the valence band plus the ion cores has a net charge of $+q$, because the
ideal crystal that included the missing electron had to be charge neutral. Thus the valence-band system has a positive charge, positive momentum, and carries a positive current.
These are the properties of a positively-charged quasi-particle moving with a postive velocity.  Moreover, the total energy of the valence-band system has increased by 
$p^2/2|m^*_v|$ because the electron at ${\bf k}=-0.2$ has effectively been raised to the valence-band edge, which requires the addition of precisely that amount of 
energy to the system.  Again, we can consider this to be the kinetic energy of a positivley-charged, positive effective-mass quasi-particle.  Thus we have found no contradictions
between the behavior of a missing electron in the valence band, and the assumed behavior of the hole quasiparticle.  Also note that at any instant one identifiable valence electron 
is responsible for the current and momentum, but the identity of that electron changes continuously under an applied field.  Hence the designation ``quasi-particle'' is appropriate.

\section{Impurity States}

The notion of the direct subtraction of the ideal host crystal immediately clarifies the behavior of substitutional impurities (also known as dopants).  These are typically described
in terms of the number of valence-shell electrons that they contribute, either more or less than the host atoms.  Expressed in terms of subtraction of the host crystal, the dopants 
are immediately seen as ``effective atoms'' that can have positive or negative effective nuclear charges (or more precisely ionic-core charges), and positive or negative electron
occupation numbers, as illustrated in Fig.\ \ref{fig:MesoDifference}.  Dopant atoms from adjacent columns of the periodic table are known as ``hydrogenic'' dopants because we can 
describe the bound states that they produce by invoking the Coulomb-potential solution of Schr\"odinger's equation, but using the mesoscopic form of that equation (\ref{eqn:effMassSchEq}).\cite{Slater1949}
In a semiconductor the atomic radius and ground-state energy are rescaled to:
\begin{gather}
 a = \frac{K_s}{|m^*_b|} a_\text{B} ,\\
 E_\text{impurity} = E_b - \frac{m^*_b}{K_s^2} R, \label{eqn:hydrogenicImpE}
\end{gather}
where $a_\text{B}$ is the Bohr radius and $R$ is the Rydberg energy.  Given the magnitude of the materials constants noted above, the orbital radius of the impurity bound state is
of the order of 10 nm, and the binding energy (relative to the band edge) is of the order of 10 meV. The quasiparticles do not remain bound in such states at normal temperatures and thus doping
with hydrogenic impurities is an effective way to introduce free charge carriers. Equation (\ref{eqn:hydrogenicImpE}) is quite accurate in cases where a single 
parabolic energy band is a good approximation.  Impurities which are two columns removed from the host atom generally show two ionization energies, similar to those of helium but on 
the reduced energy scale.

The mesoscopic transformation also helps us explain the behavior of transition-metal impurities in semiconductors, even though their states are not at all those of the ``effective atoms''
described above.  To understand the behavior of the transition metals, we must recognize that the ideal crystal which we subtract from the nonideal system contains only valence-band states 
that are derived
from s and p orbitals.  A filled d shell will be treated as part of the ionic core, but empty or partially-filled d shells whose energies lie in proximity to the those of the valence s and p states
will not be subtracted out of the problem, and thus they remain as a part of the residual system.  The d orbital energies will of course be perturbed by the chemical environment of the crystal,
but if they occur within the energy-band gap, they can appear at essentially any energy, and they will form the localized states known as ``deep levels.'' Because they are fundamentally the orbitals
of the free atom, they will have a spatial extent of the order of 0.1 nm, unlike the much more extended hydrogenic states.  Their distribution in momentum space will therefore span nearly the
entire Brillouin zone of allowed momenta, and this points to the significance of such states: They can act as intermediate states in electron-hole recombination process, effectively breaking
the momentum-conservation restriction.  This is their main technological significance, increasing the rate of recombination.  Often this is a parasitic process that we desire to minimize, but
in some cases it is desirable and can be enhanced by the deliberate introduction of transition-metal impurities.

\section{Application to Semiconductor Heterostructures}

Until the mid-1970s the picture provided by the mesoscopic transformation was primarily of use as an intermediate step in the development of the purely classical picture of semiconductors as hosts
for Newtonain quasi-particles.  But then, the development of technologies for making high-quality semiconductor \emph{heterojunctions}\cite{Kroemer1996} between chemically different materials led to the demonstration of
size-quantization effects in such structures,\cite{Dingle1974} and the \emph{quantum well} became a common technological component.  If the material composition changes abruptly the energy bandgap 
will also change abruptly, and there will necessarily be abrupt shifts
in the band-edge energies.  Because those energies are major components of the scalar potentials in equations (\ref{eqn:effMassSchEq}), the resulting Schr\"odinger problems can very closely approximate
the piecewise-constant potentials that had long appeared in textbooks on quantum mechanics.  The mesoscopic picture was then and remains today the most convenient way to evaluate the quantum 
states of such structures.  But, the presence of more than one type of ideal crystal poses complications for the formulation of the mesoscopic transformation.  

To preserve simplicity in the mesoscopic model, we prefer a transformation which subtracts out the appropriate semiconductor composition in each region of the heterostructure.
But to achieve this we need to more closely examine the technical details of the widely-used mesoscopic transformations.  

The mesoscopic transformation always consists of a unitary transformation onto a basis that represents the ideal-crystal energy-band structure, followed by a projection down onto a small number of bands.
The Wannier-Slater\cite{Slater1949} formulation expands the microscopic wavefunction in terms of Wannier functions centered on each lattice point.  The coefficients of this expansion form the mesoscopic wavefunction, and 
thus the exact formulation is spatially discrete like a tight-binding model; the continuum effective-mass Schr\"odinger differential equation is an approximation to the exact discrete difference equation.

The Luttinger-Kohn\cite{Luttinger1955} approach defines the mesoscopic wavefunction as a slowly-varying envelope function, which yields the microscopic wavefunction when multiplied by a zone-center
Bloch function.  The Fourier spectrum of the envelope function is limited to the first Brillouin zone, and any function with this property can be exactly specified by sampling its value at each lattice 
point.  Thus the Luttinger-Kohn Hamiltonian can be formulated, in principle, as a discrete matrix of the same form as the Wannier-Slater Hamiltonian. (The discrete formulation makes a fleeting 
appearance in that paper, but the authors treated it as an approximation, not recognizing that the localized functions they defined are mutually orthogonal.)

Because these transformations can be expressed in terms of localized basis functions, it would appear to be straightforward to construct a mesoscopic transformation for a hererostructure by 
using a localized basis that is derived from the correct material at each lattice point.  Burt\cite{Burt1988} noted that this does not generally produce a unitary transformation because the basis 
states from different materials are not necessarily orthogonal.  The common practice in this field has been to neglect such considerations and assume that a suitable transformation can be 
defined in which the bulk effective-mass Schr\"odinger equation is valid right up to a heteojunction.  On the whole such an approximation is well suported by agreement between experimental observations and
theoretical calculations; there is no strong evidence that the electronic structure is significantly perturbed by the heterojunction beyond the atomic layers adjacent to the junction.    

When we include the simplest models for both heterostructure and band structure the effective-mass Schr\"odinger equation becomes (in one dimension):
\begin{multline} \label{eqn:effMassSchEq2}
  i\hbar\frac{\partial\Psi_b(z)}{\partial t} = \\ - \frac{\hbar^2 }{2} \frac{\partial}{\partial z}m^*(z) \frac{\partial}{\partial z} \Psi_b + \left[E_b(z) + V_\text{ext}(z) \right] \Psi_b(z) .
\end{multline}
We must write the kinetic-energy term in a Sturm-Louville form\cite{BenDanielDuke1966} to preserve the Hermiticity of the Hamiltonian.  This is an adequate approximation when we can neglect 
energy-band nonparabolicity and the coupling of unlike bands across the heterojunction.  It leads to the continuity of $(1/m^*)(\partial\Psi/\partial z)$ as the interface matching condition for
the wavefunction.

A Hermitian mesoscopic Hamiltonian will determine the form of both interface matching conditions and of the current-density expression.  If these are derived from the same Hamiltonian they will necessarily
be self-consistent.  However, for more complicated Hamiltonians, one must actually derive the current density by deriving the continuity equation.  Generalizations of the Green identity required 
for such a derivation exist for all orders of derivative, for multi-component wavefunctions, and for spatially discrete formulations.\cite{Frensley1995}  There is a widespread misconception that the
current density can be derived from the Heisenberg equation for $dz/dt$.  While this does give the correct answer for the motion of the centroid of a wavepacket, it does not give a current density 
which obeys the expected continuity equation in discrete formulations. 

\section{The Position-Dependent Band Edge}

Finally, let us point out the role of the mesoscopic transformation in defining the notion that has underlain all of the preceeding discussion, and indeed all of semiconductor device physics:
the concept of a \emph{position-dependent band-edge}.  In principle such a concept runs into difficulties with the uncertainty principle, since energy bands are only well-defined in momentum space.
In practice, we will see that the only useful definition of such a quantity is as that potential
which appears in an appropriate effective-mass Schr\"odinger equation.  
\begin{figure}[t!b!]
\centering
\epsfig{file=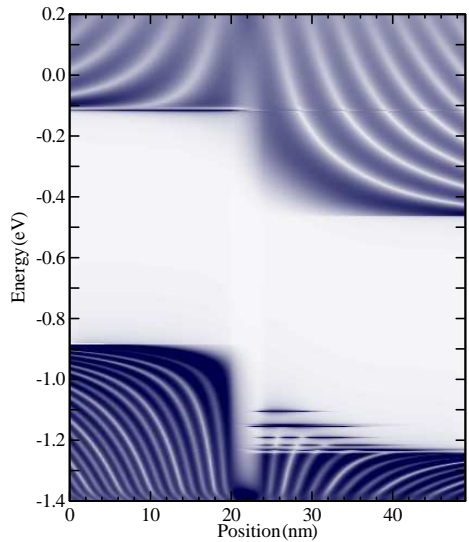, width=2.4in}
\caption{The local density of states (shown here in grayscale) is the best information that a microscopic electronic-struture computation can typically provide.}
\label{fig:denSt1}
\end{figure}
\begin{figure}[b!]
\centering
\epsfig{file=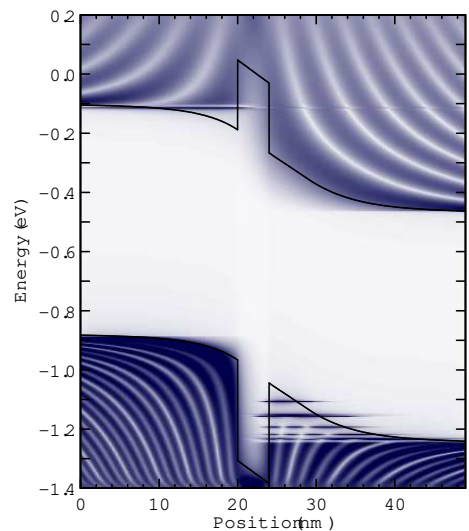, width=2.4in}
\caption{The mesoscopic picture provides a way to define position-dependent band edges which Fig.\ \ref{fig:denSt1} lacks. This makes the physics of the structure far more comprehensible.}
\label{fig:denSt2}
\end{figure}
The local band edge is not a directly observable quantity.  Consider the problem of trying to infer the energy-band profile 
from a microscopic electronic-structure computation, such as are done with density-functional methods.  Such a computation produces a set of wavefunctions and their energies, and with this information
one can construct a local density of states, such as that illustrated in Fig.\ \ref{fig:denSt1} (though this particular example was generated using a mesoscopic model).  
One can discern that there is
an energy barrier near the center that extends into both conduction and valence band, and that an overall bias voltage has been applied, due to the offset of the energy gap.

When the mesoscopic potentials are drawn on the same plot as shown in Fig.\ \ref{fig:denSt2}, the situation becomes a great deal more obvious.  One can  readily understand how the how the 
confined states are formed after viewing the potential plot.  While it is conventional to draw the band edges as continuous curves, in fact the mesoscopic picture cannot be resolved within the
primitive unit cell.  Therefore, we should really plot these quantities as discrete points.  It is the inherent graininess of the mesoscopic representation that accommodates the restrictions of
the uncertainty principle.

\section{Summary}

The concept of the mesoscopic trasformation is simple enough to be grasped and used by those who are just begining the study of semiconductor physics, even though the rigorous demonstration of 
its validity is a far more advanced task.  A distinguished precedent for the employment of such a result can be found in Richard Feynman's use of the Heaviside radiation formula to teach the 
fundamental concepts of electromagnetic radiation and of radiating systems \emph{before} developing Maxwell's equations.\cite{FeynmanVol1}  This allowed him to explain not only simple properties like interference and
polarization, but also such advanced phenomena as bremsstrahlung and synchrotron radiation.  Similarly, the mesoscopic transformation makes it easy to grasp the basic properties of electrons, holes and
dopants, but also offers insight into more advanced topics like deep-level impurities and size quantization in heterostructures.  Neither the Heaviside formula nor the mesoscopic transformation is necessarily the optimum approach
if highly accurate quantitative predictions are the goal, but each of them helps one develop an ability to identify the most likely behavior of a physical system without recourse to detailed computation.
And such an ability is the operational definition of physical intuition.

\bibliographystyle{apsrev4-1}
\bibliography{FrensleyMesoscopic}

\end{document}